\documentstyle[12pt]{article}

\begin{document}
\title{Extended Voros product in the coherent states framework}
\author{{\bf M. Daoud} \\
\\
Abdus Salam International Centre for Theoretical Physics \\ Trieste, Italy\\
\\and\\
LPMC , Department of physics, University Ibn Zohr\\  Agadir, Morocco}
\maketitle

\begin{abstract}
 Using coherent states of the Weyl-Heisenberg algebra $h_N$ , extended Voros products and Moyal brakets are derived. The covariance of Voros product under canonical transformations is discussed. Star product related to Barut-Girardello coherent states of the Lie algebra $su(1,1)$ is also considered. The star eigenvalue problem of singular harmonic osillator is investigated.
\end{abstract}

\vfill
\newpage 

\section{Introduction}
Deformation quantization is an idea to quantize classical mechanical systems without using operator theory but by deforming a Poisson algebra on a manifold into a noncommutative algebra. In classical mechanics, observables are smooth functions on phase space, which constitute a Poisson algebra, wile in quantum mechanics, the observables constitute a non-commutative associative algebra. In 1949, Moyal introduced a new bracket for functions on the classical phase space that replaces the Poisson one in the quantization procedure [1]. This bracket is closely related to Weyl's correspondence rule between classical and quantum observables [2]. The new Lie algebra associated with this bracket is a deformation of the Poisson algebra. In recent times , deformation quantization has been explored in several contexts : in the strings theory approach to non-commutative geometry [3], matrix model [4], the non-commutative Yang-Mills theories [5] and non-commutative gauge theories [6]. The apparence of noncommutatity in high energy physics helped to revive the deformation quantization technique which was elucited further in [7]. Another star product due to Grosse and Presnajder [8], using generalized coherent states [9], leads to  a quantization scheme analogue to Berezin one [10]. Through this article, we will rely on the concept of coherent states which have important application building a bridge between the classical and the quantum worlds views. In other hand, any set of coherent states satisfy two important properties: continuity and identity to unity. The resolution to unity and non-orthogonality are key ingredients to formulate the Voros product in the coherent states framework (Moyal and Voros products are equivalents) as it has been shown by Stern and al [11](see also [12]). In the same spirit, coherent states of quantum system with nonlinear spectrum has been considered to find new kinds of star products [13] and the $su(2)$ coherent states has been nicely used to introduce a new star product on the fuzzy sphere [14]. \\
Usually the coherent states of the Weyl-Heisenberg algebra $h_N$
are constructed by taking the vacuum as reference state and they are eigenvectors of annihilation operators (standard coherent states)
[15]. However, they
can be also defined by acting the unitary displacement operator
on arbitrary element of the representation space of $h_N$
algebra leading to the so-called generalized coherent states [9]. The first main of this work is to point out to extend
the standard Voros star product using these coherent states. In
the next section, we first review some basic facts about coherent
states of $h_N$ algebra [9] (See also the references quoted in [16-17]), and the associated Voros product. Using
coherent states constructed from arbitrary reference state, we
generalize the standard Voros product.  In the best of our
knowledge, this extension has been not discussed previously. As
illustration, we consider Landau problem. The third section
concern the unitary transformation in the noncommutative space
spanned by the variables associated to generators of the algebra
$h_N$. It will be shown that the covariance is violated under
arbitrary unitary transformations and the particular case in
which the covariance is assured corresponds to the canonical
transformations of $h_N$ elements. We discuss also the link
between squeezing of coherent states and canonical transformation
in the noncommutative space generated by eigenvalues variables
associated to generators of Weyl-Heisenberg algebra. The last
section is devoted to star product defined by mean of $su(1,1)$
coherent states. As application of this new star product, we
investigate the star analogue of eigenvalue problem of the the
so-called  singular harmonic oscillator. The quantization of the
quantum mechanics of this system is different from Berezin one
[10]. Concluding remarks close this paper.
\section{\bf Coherent states and star products}
In this section, we give a review of  standard coherent states
[15] and derive the  associative Voros star product using
these states. We first consider $n = 2N$ canonical $a_{i}^+$
(boson creation) and $a_{i}^-$ (boson destruction) operators, $i
= 1, 2,...,N $. The set of operators $\{a_{i}^+,a_{i}^-\}$ and
the unity close the Weyl-Heisenberg algebra $h_N$ which describe
a bosonic system with $N$ degrees of freedom. The operators
$a_{i}^+$ and $a_{i}^-$ act in the Hilbert space ${\cal H}_b =
{\cal H}_1\otimes{\cal H}_2\otimes ....\otimes{\cal H}_N $
 generated by the states

$|n_1, n_2,...,n_N\rangle = |n_1\rangle\otimes|n_2\rangle\otimes...\otimes|n_N\rangle$
where $n_i$ are nonnegative integers. The coherent states for $h_N$ algebra are
defined as eigenstates of the operators $a_{i}^-$ with eigenvalues $z_i$:\\
\begin{equation}
\vert\vec{z} \rangle  \equiv |z_1, z_2,...,z_N\rangle = e^{-\sum_{i=1}^N{\vert z_{i} \vert}^2}\sum_{n_{i}=0}^\infty \frac{z_{1}^{n_1}z_{2}^{n_2}...z_{N}^{n_N}}{\sqrt{n_{1}!n_{2}!...n_{N}!}} \vert n_1, n_2,...,n_N\rangle
\end{equation}
The states $\vert\vec{z} \rangle$ can be used, following the method presented in [10], to introduce an associative star product.\\
\hspace*{0.5cm}To every operator $O_1$ acting on the Hilbert space ${\cal H}_b$
\begin{equation}
O_1 = \sum_{\vec n , \vec m} (O_1)_{\vec n , \vec m} {\hskip 0.5cm} ({a_{1}^+})^{m_1}...({a_{N}^+})^{m_N}({a_{1}^-})^{n_1}...({a_{N}^-})^{n_N}
\end{equation}
where $\vec m = ( m_1,...,m_N )$ and $\vec n = ( n_1,...,n_N )$, one can associate a function ${\cal O}_{1}( \vec {\bar z} , \vec z )$ according
\begin{equation}
{\cal O}_{1}( \vec {\bar z} , \vec z ) = \langle \vec z \vert O_{1} \vert\vec z \rangle
                                = \sum_{\vec n , \vec m} (O_1)_{\vec n , \vec m} {\hskip 0.5cm} ({\bar z_{1}})^{m_1}...({\bar z_{N}})^{m_N}({z_{1}})^{n_1}...({z_{N}})^{n_N}
\end{equation}
An associative star product of two functions ${\cal O}_{1}( \vec {\bar z} , \vec z )$ and ${\cal O}_{2}( \vec {\bar z} , \vec z )$ is then defined by
\begin{equation}
{\cal O}_{1}( \vec {\bar z} , \vec z ) \star {\cal O}_{2}( \vec {\bar z} , \vec z ) = \langle \vec z \vert O_{1} O_{2} \vert\vec z \rangle
\end{equation}
The associativity of the this star product originates from associativity of the Weyl-Heisenberg algebra $h_N$.
Since the coherent states $\vert\vec z \rangle$ are eigenstates of the operators $a_{i}^-$ with the eigenvalues $z_{i}$, it is easy to verify that
\begin{equation}
\bar z_{i} \star z_{j} = \bar z_{i} z_{j}
\end{equation}
\begin{equation}
\bar z_{i} \star \bar z_{j} = \bar z_{i} \bar z_{j} = \bar z_{j} \star \bar z_{i}
\end{equation}
\begin{equation}
z_{i} \star z_{j} =  z_{i} z_{j} =  z_{j} \star  z_{i}
\end{equation}
\begin{equation}
z_{j} \star \bar z_{i} = \delta_{ij} + z_{j} \bar  z_{i}
\end{equation}
The Moyal Brackets are given by
\begin{equation}
\{z_{j}, \bar z_{i}\}_{\star} = z_{j} \star \bar z_{i} - \bar z_{i} \star z_{j} = \delta_{ij}
\end{equation}
reflecting that the structure relations of $h_N$ algebra  are preserved in the star quantization.
 Note that the relations (5-8) constitute the minimal set of relations to compute the $\star$-product of any two arbitrary functions ${\cal O}_{1}( \vec {\bar z} , \vec z )$ and $ {\cal O}_{2}( \vec {\bar z} , \vec z )$. Note also that the coherent states $\vert \vec z \rangle$ correspond to unitary action on the ground state $\vert 0 ,0,...,0\rangle$ of the multimode bosonic system under consideration. As mentionned in the introduction, general sets of coherent states  can be constructed by acting the unitary operator
\begin{equation}
D(\vec {z}) = e^{\sum_{i=1}^{N}(z_{i}a_{i}^{+} - \bar z_{i} a_{i}^{-})}
\end{equation}
on arbitrary reference state $ \vert k_{1},k_{2},...,k_{N}\rangle $ and can used to define a new star products,
extending the Voros one. The action of the operator $D(\vec {z})$ on the reference state $\vert k_{1},k_{2},...,k_{N}\rangle \equiv \vert \vec {k}\rangle $ :
\begin{equation}
\vert \vec {z},\vec {k}\rangle = \otimes_{i=1}^{N} D(z_{i}) \vert k_{i}\rangle  \equiv  \otimes_{i=1}^{N} \vert z_{i},k_{i}\rangle
\end{equation}
where the vectors $ \vert z_{i},k_{i}\rangle$ are expressed as follows
\begin{equation}
\vert z_{i},k_{i}\rangle = e^{-\frac{1}{2}z_i\bar z_i}\bigg(
\sum_{l_{i}\le k_{i}} \sqrt{\frac{l_{i}!}{k_{i}!}} (-\bar z_{i})^{k_{i}- l_{i}}
L_{l_{i}}^{k_{i}- l_{i}}(\vert z_{i}\vert^{2})\vert l_{i}\rangle
+\sum_{l_{i}\ge k_{i}}  \sqrt{\frac{k_{i}!}{l_{i}!}}  (z_{i})^{l_{i}- k_{i}}
L_{k_{i}}^{l_{i}- k_{i}}(\vert z_{i}\vert^{2})\vert l_{i}\rangle\bigg)
\end{equation}
in term of the Laguerre polynomials $L_{n}^{\alpha}$.\\
The states (11) resolve the unity operator for any $\vec {k}= (k_{1},k_{2},...,k_{N})\in {\bf{N}}^{N}$ in respect to the measure $\frac{ d^{2}\vec {z}}{\pi^N}$.\\
The associative product of two functions ${\cal O}_{1}( \vec {\bar z} , \vec z )$ and ${\cal O}_{2}( \vec {\bar z} , \vec z)$ is defined, in this case, by
\begin{equation}
{\cal O}_{1}( \vec {\bar z} , \vec z ) \star_{k} {\cal O}_{2}( \vec {\bar z} , \vec z ) = \int d^{2}\vec {z'}
\langle \vec {z},\vec {k}\vert O_{1}\vert \vec {z'},\vec {k}\rangle
\langle \vec {z'},\vec {k}\vert O_{2}\vert \vec {z},\vec {k}\rangle
\end{equation}
Noticing that
\begin{equation}
\prod_{i=1}^{N} e^{-z_{i}\frac{\partial}{\partial z'_{i}}}e^{z'_{i}\frac{\partial}{\partial z_{i}}}{\cal O}_{1}( \vec {\bar z} , \vec z ) = \frac{\langle \vec {z},\vec {k}\vert O_{1}\vert \vec {z'},\vec {k}\rangle}{\langle \vec {z},\vec {k}\vert \vec {z'},\vec {k}\rangle}
\end{equation}
and
\begin{equation}
\prod_{i=1}^{N} e^{-\bar {z_{i}}\frac{\partial}{\partial\bar  {z'_{i}}}}e^{\bar {z'_{i}}\frac{\partial}{\partial \bar {z_{i}}}}{\cal O}_{2}( \vec {\bar z} , \vec z ) = \frac{\langle \vec {z'},\vec {k}\vert O_{2}\vert \vec {z},\vec {k}\rangle}{\langle \vec {z'},\vec {k}\vert \vec {z},\vec {k}\rangle},
\end{equation}
one can write the $\star_k$-product as
\begin{equation}
\star_{k} = \int d^{2}\vec {z} :e^{\sum_{i=1}^{N}\frac{\partial}{\partial z_{i}}(z'_{i}-z_{i})}:
\vert \langle \vec {z},\vec {k}\vert \vec {z'},\vec {k}\rangle \vert^{2}
:e^{\sum_{i=1}^{N}(\bar z'_{i}-\bar z_{i})\frac{\partial}{\partial \bar z_{i}}}:
\end{equation}
where the symbol : : stand for an ordered exponential. Clearly the star product is completely determined once one known the overlapping, between two coherent states (11), which is given by
\begin{equation}
\vert \langle \vec {z},\vec {k}\vert \vec {z'},\vec {k}\rangle \vert^{2} =  \prod_{i=1}^{N} e^{-\vert z'_{i}- z_{i}\vert^{2}}\big( L_{k_{i}}^{0}(\vert z'_{i}- z_{i}\vert^{2})\big)^2
\end{equation}
Substituting (17) in (16) and with a simple change of variables, one get
\begin{equation}
\star_{k} = \prod_{i=1}^{N}\star_{k_{i}}
\end{equation}
where
\begin{equation}
\star_{k_{i}} = \sum_{p=0}^{\infty} I_{k_{i}p}\overleftarrow{\frac{\partial ^p}{\partial z_{i}^p}}\overrightarrow{\frac{\partial ^p}{\partial \bar z_{i}^p}}
\end{equation}
The coefficients $ I_{k_{i}p}$ occuring in the last equation are
given by
\begin{equation}
I_{k_{i}p} = \sum_{j,j'=0}^{k_{i}}\frac{(p+j+j')!}{(p!)^2}(-)^{j+j'}{k_{i}\choose k_{i}-j}{k_{i}\choose k_{i}-j'}
\end{equation}
In the particular case $k_{1}=k_{2}=...=k_{N}=0$, we have $I_{0p}=\frac{1}{p!}$ and we recover the Voros star product
\begin{equation}
\star = \star_{0} = e^{\sum_{i=1}^N\overleftarrow{\frac{\partial ^p}{\partial z_{i}^p}}\overrightarrow{\frac{\partial ^p}{\partial \bar z_{i}^p}}}.
\end{equation}
So, it becomes clear that the coherent states (11) leads to a new star product generalizing the Voros one. To compute the $\star_{k}$-product, of two arbitrary functions, the desired relations are
\begin{equation}
\bar z_{i} \star_k z_{j} = I_{kN} \bar z_{i} z_{j}
\end{equation}
\begin{equation}
\bar z_{i} \star_k \bar z_{j} = I_{kN} \bar z_{i} \bar z_{j} = \bar z_{j} \star_k \bar z_{i}
\end{equation}
\begin{equation}
z_{i} \star_k z_{j} =  I_{kN} z_{i} z_{j} =  z_{j} \star_k z_{i}
\end{equation}
\begin{equation}
z_{j} \star_k \bar z_{i} = I_{kN}[ \delta_{ij} \frac{I_{k_i1}}{I_{k_i0}} + z_{j} \bar  z_{i}]
\end{equation}
where the c-number $ I_{kN}$ is defined by $ I_{kN} = \prod_{i=1}^N  I_{k_i0}$. In this case, the extended Moyal brackets are given by
\begin{equation}
\{z_{j}, \bar z_{i}\}_{\star_k} = z_{j} \star_k \bar z_{i} - \bar z_{i} \star_k z_{j} = I_{kN}\frac{I_{k_i1}}{I_{k_i0}} \delta_{ij}
\end{equation}
Note that for $k_1=k_2=...=k_N=0$, the relations (26) reduces to ones given by (9) and we recover the usual Moyal brackets.\\
The physical motivation of the extension of the Voros product can be found in the well known Landau problem. More precisely, the higher landau levels quantum mechanics can be equivalently formulated in a noncommuative setting involving the extended Voros products and to each Landau level $k$ a class of Voros product $\star_k$ can be associated. Before this simple illustration, recall that the Landau spectrum is made of degenerate Landau levels $E_k = 2k+1$ ($k \ge 0$) with $k$th Landau level eigenstates labelled by the radial/orbital quantum numbers $k$ , $l \ge 0$ (analytic) and $k+l$ , $-k \le l \le 0$ (anti-analytic). There are, in a given Landau level, an infinite number of analytic eigenstates
\begin{equation}
\Phi_{k,l}(z,\bar z) = e^{-\frac{1}{2}|z|^2} z^l L_{k}^{l}(|z|^2) {\hskip 1cm} l \ge 0
\end{equation}
and a finite number of anti-analytical eigenstates
\begin{equation}
\Phi_{k,k+l}(z,\bar z) = e^{-\frac{1}{2}|z|^2} {\bar z}^{-l} L_{k+l}^{-l}(|z|^2) {\hskip 1cm} -k \le l < 0
\end{equation}
It is interesting to note that the analytic and anti-analytic functions of the $k$th Landau level can be written also as
\begin{equation}
\Phi_{k,l}(z,\bar z) = (-)^l \sqrt{\frac{(k+l)!}{k!}}\langle k+l \vert z , k \rangle {\hskip 1cm} l \ge 0
\end{equation}
and
\begin{equation}
\Phi_{k,k+l}(z,\bar z) = \sqrt{\frac{k!}{(k+l)!}}\langle k+l \vert z , k \rangle {\hskip 1cm} -k \le l < 0
\end{equation}
It is clear that the functions $\Phi_{k,l}(z,\bar z)$ (resp. $\Phi_{k,k+l}(z,\bar z)$) corresponds to analytic (resp. anti-analytic) representations of the coherent state $\vert z , k \rangle$ (Equation (11) for one bosonic degree of freedom) constructed from the fiducial vector $\vert k \rangle$ (eigenstate of the Landau Hamiltonian). In view of the above considerations on the extension of the Voros product, it results that in a given Landau level $k$, the noncommutativity can be introduced through $\star_k$ product between the analytic representations of degenerate states. Of course, this question requires more analysis which will be considered in another work.
\section{Canonical covariance }
\hspace*{0.5cm}Let us start by examining the covariance in the noncommutative space, endowed with  Voros product, under unitary transformations. For simplicity reasons, we restrict ourself to one bosonic degree of freedom. Let $O$ be an operator and $U = e^{\Lambda}$ $(\Lambda^{+} + \Lambda = 0)$ a one or multi-parameter unitary transformation of the operator $O$ in the operator space. First, we will show that the unitary transformations in the operator space has unique representation in the noncommutative space generated by the variables $z$ and $\bar z$. In other words, we will show that the function ${\cal O}'(z,\bar z)$ associated to the operator $O'= U^+ O U$ is given by
\begin{equation}
{\cal O}'(z,\bar z) = e^{-{\cal D}_{\lambda}}{\cal O}(z,\bar z)
\end{equation}
where the function ${\cal O}(z,\bar z) = \langle z \vert O \vert z \rangle$ and ${\cal D}_{\lambda}$ acts as
\begin{equation}
{\cal D}_{\lambda}
{\cal O}(z,\bar z) = \lambda(z,\bar z) \star {\cal O}(z,\bar z) - {\cal O}(z,\bar z) \star \lambda(z,\bar z)
\end{equation}
with $\lambda(z,\bar z) =  \langle z \vert \Lambda \vert z \rangle $.\\
As we mentioned in the previous section, any operator $O$ in the
operator algebra can be expanded in terms of creation and
annihilation operators $a^+$ and $a^-$
\begin{equation}
O = \sum_{m,n} O_{m,n} (a^+)^{m}(a^-)^{n}
\end{equation}
The proof of the relation (31) is immediate. Indeed, expanding the operator $O'$ as
\begin{equation}
O' = \sum_{k=0}^{\infty} \frac{(-1)^k}{k!} \bigg[ \Lambda\big[ \Lambda ,...,[ \Lambda , O ]...\big]\bigg]
\end{equation}
and using the star calculus in the coherent states scheme, it is easily seen that the function ${\cal O}'(z,\bar z)$ is given by
\begin{equation}
{\cal O}'(z,\bar z) = \sum_{k=0}^{\infty} \frac{(-1)^k}{k!} \bigg\{ \lambda(z,\bar z)\big\{ \lambda(z,\bar z) ,...,\{ \lambda(z,\bar z) , {\cal O}(z,\bar z) \}_{\star}...\big\}_{\star}\bigg\}_{\star}
\end{equation}
A result which can be also written in the compact form as in (31) where ${\cal D}_{\lambda}$ is defined through Eq(32). Therefore, each unitary transformation in the operator space has unique representation in the noncommutative space and there a correspondance between unitary transformations in the operator space  and ones given by (31). Note also that we have the following identities
\begin{equation}
e_{\star}^{- \lambda}\star e_{\star}^{\lambda }= e_{\star}^{\lambda} \star e_{\star}^{-\lambda} = 1
\end{equation}
where the star exponential is defined as
\begin{equation}
e_{\star}^{X} = \sum_{k=0}^{\infty}\frac{1}{k!}(X)^{\star k} = \sum_{k=0}^{\infty}\frac{1}{k!} X \star X \star ...\star X
\end{equation}
Remark that if we consider two transformations acting on the operator space $O_1 = e^{\Lambda_{1}}$ and   $O_2 = e^{\Lambda_{2}}$, one has
\begin{equation}
[ {\cal D}_{\lambda_1} , {\cal D}_{\lambda_2}] = {\cal D}_{\{\lambda_1 , \lambda_2\}_{\star}}
\end{equation}
where the functions $\lambda_1$ and $\lambda_2$ are ones associated with anti-hermitians operators $\Lambda_1$ and $\Lambda_2$, respectively.\\
Using the harmonic coherent states, the function associated to the operator $O$ (Eq.(33)) is
\begin{equation}
{\cal O}( z , \bar z) = \sum_{mn} O_{mn} {\bar z}^m z^n ,
\end{equation}
and one corresponding to the operator $ O' = U^+ O U $ is given by
\begin{equation}
{\cal O}'( z , \bar z) = \langle z \vert e^{- \Lambda} O  e^{+ \Lambda}\vert z \rangle
\end{equation}
Thanks to the definition (4), the function ${\cal O}'( z , \bar z)$ can be written  as
\begin{equation}
{\cal O}'( z , \bar z) = \langle z \vert e^{- \Lambda} \vert z \rangle \star
\langle z \vert  O  \vert z \rangle \star
\langle z \vert  e^{+ \Lambda}\vert z \rangle
\end{equation}
and combining the definitions (4), (32) and (41), one has
\begin{equation}
{\cal O}'( z , \bar z) = e_{\star}^{-\lambda (z,\bar z)} \star {\cal O}( z , \bar z) \star e_{\star}^{+\lambda (z,\bar z)}
\end{equation}
Alternatively, It can be expressed as follows
\begin{equation}
{\cal O}'( z , \bar z) = \sum_{mn} O_{mn} \langle z \vert e^{- \Lambda} (a^+)^m  e^{+ \Lambda}\vert z \rangle \star \langle z \vert e^{- \Lambda} (a^-)^n e^{+ \Lambda}\vert z \rangle
\end{equation}
\begin{equation}
{\cal O}'( z , \bar z) =  \sum_{mn} O_{mn} {\bar Z}^{\star m} \star Z^{\star n}
\end{equation}
where the new variables $Z$ and $\bar Z$ are defined by
\begin{equation}
Z =  e^{-{\cal D}_{\lambda}} z {\hskip 2cm} \bar Z =  e^{-{\cal D}_{\lambda}} \bar z
\end{equation}
It is evident that
\begin{equation}
{\cal O}'( z , \bar z) \ne {\cal O}( Z , \bar Z)
\end{equation}
in general. Thus, the covariance under arbitrary unitary transformation is violated. The considerations presented here can be extended to multi-dimensional noncommutative space in a straightfoward way.\\
One exceptional case in which the  equality in (46) holds  can be identified corresponds to the group of linear canonical transformations. To examine this exceptional case, let us first recall the canonical transformations for a bosonic system with a finite number of degrees of freedom. We shall consider the transformations preserving the commutations relations of $h_N$ algebra
\begin{equation}
A_i^- = T^+ a_i^- T.
\end{equation}
The operator $T$ is given by
\begin{equation}
T = \exp \big ( \frac{\xi_{ij}}{2}a_{i}^+a_{j}^+ -  \frac{\bar \xi_{ij}}{2}a_{i}^-a_{j}^- \big)
\end{equation}
where $\xi_{ij}$ are elements of complex symmetrical matrix that will denoted by $\Xi$. In (48),  summation over repeated indices is underlying. The operator $T$ realize a representation of the group $Sp(2N,\bf R)$. By a direct computation, one can write the transformation (47) as follows
\begin{equation}
A^{\mp} = \cosh \sqrt{\Xi^+\Xi} a^{\mp} + \frac{\Xi}{\sqrt{\Xi^+\Xi}}\sinh \sqrt{\Xi^+\Xi} a^{\pm}
\end{equation}
in a compact form in term of the complex symmetrical matrix $\Xi$.\\
In other hand, the function  corresponding to the operator $\Lambda = \frac{\xi_{ij}}{2}a_{i}^+a_{j}^+ -  \frac{\bar \xi_{ij}}{2}a_{i}^-a_{j}^-$ (Eq.(48)) takes the following form
\begin{equation}
\lambda (\vec z , \vec {\bar z}) = \frac{1}{2}\xi_{ij}\bar z_i\bar z_j - \frac{1}{2}\bar \xi_{ij} z_i z_j
\end{equation}
Using the latter expression of the function $\lambda (\vec z , \vec {\bar z})$, one can evaluate the action of the operator ${\cal D}_{\lambda}$ on canonical variables $z_k$ and $\bar z_k$. We obtain
\begin{equation}
{\cal D}_{\lambda}z_k = - \xi_{ki}\bar z_i  {\hskip 1cm} {\cal D}_{\lambda}\bar z_k = - \xi_{ki}z_{i}
\end{equation}
which can be written also in compact form as
\begin{equation}
{\cal D}_{\lambda}z = - \Xi \bar z {\hskip 1cm} {\cal D}_{\lambda}\bar z = - \Xi^+ z
\end{equation}
From the later results, one show
\begin{equation}
e^{-{\cal D}_{\lambda}} z =  \cosh \sqrt{\Xi^+\Xi} z + \frac{\Xi}{\sqrt{\Xi^+\Xi}}\sinh \sqrt{\Xi^+\Xi} \bar z = Z
\end{equation}
and
\begin{equation}
e^{-{\cal D}_{\lambda}}\bar z =  \cosh \sqrt{\Xi^+\Xi} \bar z + \frac{\Xi^+}{\sqrt{\Xi^+\Xi}}\sinh \sqrt{\Xi^+\Xi} z = \bar Z
\end{equation}
to be compared with (49). Following the prescription presented in the section 2, the coherent states associated to the algebra generated by the creation and annihilation operators $\{A_i^+ , A_i^-; i = 1,2,...,N\}$ leads to a star product of Voros type
\begin{equation}
\tilde\star =  \exp\big (\sum_{i=1}^N\overleftarrow{\frac{\partial}{\partial Z_i}} \overrightarrow{\frac{\partial}{\partial {\bar Z_i}}}\big ),
\end{equation}
and we have the following star commutation relations
\begin{equation}
\bar Z_i \tilde\star Z_j - Z_j \tilde\star \bar Z_i = \delta_{ij}{\hskip 1cm}\bar Z_i \tilde\star \bar Z_j = \bar Z_j \tilde\star \bar Z_i {\hskip 1cm} Z_i \tilde\star Z_j = Z_j \tilde\star  Z_i
\end{equation}
The new variables $Z_i$ corresponds to analytic representation of the
operators $A_i^-$. Because the transformations are canonical, the
$\star$ and $\tilde\star$-products are identical and we obtain
\begin{equation}
{\cal O}'( z , \bar z) = {\cal O}( Z , \bar Z).
\end{equation}
To close this section, some remarks are in order. First, recall that in any deformation quantization scheme the classical covariance must be assured. In our case, this covariance is guaranted when the transformations in the operator space are of the form given by (47). If one consider one bosonic degree of freedom, the operator (48) is nothing but the so-called squeezing operator. His action on the coherent states generates squeezed states which minimize also the the Heisenberg uncertainty relation but with different variances of the creation and annihilation operators. Then, it seems that the correspondence operator-function using coherent states  provides a deeper insight into the properties of quantum state that can not be seen from the analytical functions. There are a parrallelism between canonical covariance in the noncommutative space and the transformation (47) which transforms a coherent state in squeezed one in a way preserving the minimal value of the Heisenberg inequality. It will be amusing to study this parrallelism in the light of the recent results related to quantum correlations and the extended phase space discussed in [18].
\section{\bf $su(1,1)$ star product}
 In 1971, Barut and Girardello [19] defined
the coherent states of $su(1,1)$ algebra as eigenstates of the lowering operator $K_{-}$. They are given by
\begin{equation}
\vert z \rangle\  = {{\vert z \vert}^{k-1/2}\over{\big(I_{2k-1}}(2\vert z \vert)\big)^{1/2}}
\sum_{n=0}^{\infty}{z^{n}\over{\sqrt{\Gamma(n+1)\Gamma(n+2k)}}}\vert
n , k \rangle
\end{equation}
where $I_{2k-1}(2 \vert z \vert)$ is the modified Bessel function and $k$ stand for discrete series of $su(1,1)$ unitary irreducible representations. The Hilbert space is spanned by the complete orthonormal basis $\vert n, k \rangle$. Due to the completion
property , any state $\vert f \rangle$, of the Hilbert space, can
be represented by an entire function. In particular, the analytical representations of the vectors $\vert n , k \rangle$ are
\begin{equation}
{\cal F}^{k}_{n}(z)  =  \frac{z^{n}}{{\sqrt{\Gamma(n+1)\Gamma(n+2k)}}}
\end{equation}
The $su(1,1)$ elements $K_{+}$, $K_{-}$ and $K_{3}$ are realized,
in this representation, by
\begin{equation}
K_{+} = z
{\hskip 1cm}
K_{-} = z {d^{2}\over{dz^{2}}} + 2k{d\over{dz}}
{\hskip 1cm}
K_3 = z{d\over{dz}} + k
\end{equation}
Similarly to the Weyl-Heisenberg algebra, to every operator $A$ of $su(1,1)$ algebra acting on the the representation space spanned by the vectors $\vert n , k \rangle$, one can associate on the complex plane a function ${\cal A}(z , \bar z)$ as
\begin{equation}
{\cal A}(z , \bar z) = \langle z \vert A \vert z \rangle
\end{equation}
 In this respect, to element $K_-$ (resp. $K_+$) correspond the analytic function $z \to z$ (resp. anti-analytic function $\bar z \to \bar z)$. The star product, in the $su(1,1)$ case, is defined in the same manner that one given by (4). The associativity of this star product is inherited, here again, from the associativity of the usual product of the $su(1,1)$ algebra. It is simple to verify that the star product of two analytic or two anti-analytic functions reduces the usual commutative product of functions. One can also verify that
\begin{equation}
\bar z \star z = \bar z z {\hskip 2cm} z \star \bar z = z \bar z  - \Theta_k (z,\bar z)
\end{equation}
where the function $\Theta_k (z,\bar z)$ is given by
\begin{equation}
\Theta_k (z,\bar z) = 2k + \frac{\vert z \vert^2}{k} \frac{_0F_1 (2k+1,\vert z \vert^2)}{_0F_1 (2k,\vert z \vert^2)}
\end{equation}
In (63), $_0F_1$ is the well known hypergeometric function. The star product between variables $z$, $\bar z$ and the function $\Theta_k (z,\bar z)$ are given by
\begin{equation}
\bar z \star \Theta_k (z,\bar z) = \Theta_k (z,\bar z) + 2k\bar z - 2k
\end{equation}
\begin{equation}
\Theta_k (z,\bar z) \star \bar z = \Theta_k (z,\bar z) + (2k+2)\bar z - 2k
\end{equation}
\begin{equation}
\Theta_k (z,\bar z) \star z  = \Theta_k (z,\bar z) + 2kz - 2k
\end{equation}
\begin{equation}
z \star \Theta_k (z,\bar z) = \Theta_k (z,\bar z) + (2k+2)\bar z - 2k
\end{equation}
which are useful to compute the star product between any two arbitrary functions of $z$ and $\bar z$. The Moyal brackets, in the $su(1,1)$ case, are given  by
\begin{equation}
\{ z , \bar z \}_{\star} = \Theta_k (z,\bar z) {\hskip 1cm} \{ z ,  \Theta_k (z,\bar z)\}_{\star} = 2z {\hskip 1cm}\{ \bar z ,  \Theta_k (z,\bar z)\}_{\star} = - 2\bar z
\end{equation}
traducing the fact that the $su(1,1)$ structure relations are preserved in the star language. It is well established that there is a contraction procedure reducing the $su(1,1)$ algebra to Weyl-Heisenberg one [19]. Following this way, we will show that the star  commutation relations (68) can be contracted to ones corresponding to harmonic oscillator. In this order, we set
\begin{equation}
z' = \sqrt q z  {\hskip 1cm} \bar z' = \sqrt q \bar z {\hskip 1cm} \Theta'_k (z,\bar z) = q \Theta_k (z,\bar z)
\end{equation}
with $ q > 0$. The relations (68) becomes
\begin{equation}
\{ z' , \bar z' \}_{\star} = \Theta'_k ({z'\over \sqrt q},{\bar z'\over\sqrt q}) {\hskip 1cm} \{ z' ,  \Theta'_k ({z'\over \sqrt q},{\bar z'\over \sqrt q})\}_{\star} = 2qz' {\hskip 1cm}\{ \bar z' ,  \Theta'_k ({z'\over \sqrt q},{\bar z'\over \sqrt q})\}_{\star} = - 2q\bar z'
\end{equation}
In the limit $k \to \infty$ and $q \to 0$ but $2qk \to 1$, by substituting (69) in (63), one get
\begin{equation}
\Theta'_k ({z'\over \sqrt q},{\bar z'\over\sqrt q}) \to 1
\end{equation}
and the star structure relations coincides with noncommutative ones associated to harmonic oscillator. \\
To give an application of the $su(1,1)$ star product, we consider the  singular harmonic oscillator. Before going on, recall that the star analogue of eigenvalue problem for n-dimensional noncommutative harmonic oscillator has been recently considered in [20].\\
The one dimensional harmonic oscillator still integrable if we add a $x^{-2}$ potential and we have the singular harmonic oscillator
\begin{equation}
H_{cal} = a^+a^- + {1\over2}+  {{\eta}^{2}\over {x^2}}
\end{equation}
where $a^{\pm} = {1\over\sqrt{2}}(x\pm {d\over{dx}})$
are the usual creation and annihilation operators.  This one dimensional system is called also "isotonic oscillator" [21] or two-particles Calogero interaction [22]. The normalized eigenfunctions of the Hamiltonian $H_{cal}$ are
\begin{equation}
\Psi_n(x) = (-)^{n} \sqrt{{2n!}\over{\Gamma(n+e_{0})}}
L^{e_{0}-1}_n (x^{2})\exp(-{x^{2}\over{2}})
\end{equation}
where $\alpha=
{1\over2}+\sqrt{({1\over4}+2{{\eta}^{2}})}$. The eigenvalues are
given by:
\begin{equation}
H_{cal} \Psi_n(x) = e_n \Psi_n(x)
\end{equation}
where $e_n = (2n+e_{0})$ and $e_{0}=\alpha+ 1/2$. The waves functions $\Psi_n(x)$ form a basis in the Hilbert space $\cal H$ of square integrable functions on the half axis $0<x<\infty$. The raising and lowering operators are
defined by
\begin{equation}
A^{\pm} = {1\over2}((a^{\pm})^{2}-{{\eta}^{2}\over {x^2}}),
\end{equation}
and act on the eigenstates $\vert\Psi_n\rangle$  as follows
\begin{equation}
A^{+}\vert\Psi_n\rangle = \sqrt{(n+1)(n + e_{0})} \vert\Psi_{n+1}\rangle
\end{equation}
and
\begin{equation}
A^{-}\vert\Psi_n\rangle = \sqrt{n(n+e_{0}-1)} \vert\Psi_{n-1}\rangle.
\end{equation}
The operaors $A^+$ , $A^-$ and $H_{cal}$ close the $su(1,1)$ algebra. Thus, the Barut-Girardello coherent states of the singular harmonic oscillator are obtained from the expression (58) by simply setting $2k\equiv e_0 + 1$ and $\vert
n , k \rangle \equiv \vert\Psi_n\rangle $.
Every operator $A$, acting on the Hilbert space of the quantum system under consideration, can be expanded in the operator basis $\{ P_{mn} =  \vert \Psi_m \rangle \langle \Psi_n\vert \}$ and to each element $P_{mn}$ we associate the function
\begin{equation}
{\cal P}_{m,n}(z , \bar z) = \langle z \vert \Psi_m \rangle \langle \Psi_n \vert  z \rangle =
\frac{\vert z \vert^{e_0}}{I_{e_0}(2 \vert z \vert)} {\cal F}^{(e_0+1)/2}_m (\bar z) {\cal F}^{(e_0+1)/2}_n (z)
\end{equation}
satisfying the completion relation $\sum_{m=0}^{\infty} {\cal P}_{m,n}(z , \bar z) = 1$ and the orthogonality property
\begin{equation}
{\cal P}_{m,n}(z , \bar z) \star {\cal P}_{m',n'}(z , \bar z) = \delta_{m',n} {\cal P}_{m ,n'}(z , \bar z)
\end{equation}
It is easy to prove that the ${\cal P}_{0,0}(z , \bar z)$ provides a  star vacuum
\begin{equation}
z \star {\cal P}_{0,0}(z , \bar z) = {\cal P}_{0,0}(z , \bar z) \star \bar {z} = 0.
\end{equation}
We have also
\begin{equation}
z \star {\cal P}_{m,n}(z , \bar z) = \sqrt{m(m+e_{0}-1)}{\cal P}_{m-1,n}(z , \bar z)
\end{equation}
\begin{equation}
\bar {z} \star {\cal P}_{m,n}(z , \bar z) = \sqrt{(m+1)(m + e_{0})}{\cal P}_{m+1,n}(z , \bar z)
\end{equation}
\begin{equation}
{\cal P}_{m,n}(z , \bar z) \star z = \sqrt{(n+1)(n + e_{0})}{\cal P}_{m,n+1}(z , \bar z)
\end{equation}
\begin{equation}
{\cal P}_{m,n}(z , \bar z) \star \bar {z} = \sqrt{n(n+e_{0}-1)}{\cal P}_{m,n-1}(z , \bar z)
\end{equation}
The function ${\cal H}_{cla}$ associated to the Hamiltonian $H_{cal}$ is given
\begin{equation}
{\cal H}_{cla} = \langle z \vert H \vert z \rangle = \Theta_{\frac{e_0+1}{2}} (z,\bar z)
\end{equation}
where the function $\Theta$ is defined by Eq(63) and one show that
\begin{equation}
{\cal H}_{cla} \star {\cal P}_{m,n}(z , \bar z) = e_m {\cal P}_{m,n}(z , \bar z)
\end{equation}
\begin{equation}
{\cal P}_{m,n}(z , \bar z) \star {\cal H}_{cla} = e_n {\cal P}_{m,n}(z , \bar z)
\end{equation}
The functions ${\cal P}_{n,n}(z , \bar z)$ are the star eigenstates of ${\cal H}_{cla}$ satisfying ${\cal H}_{cla} \star {\cal P}_{n,n}(z , \bar z) = {\cal P}_{n,n}(z , \bar z) \star {\cal H}_{cla}$ and they are real. Finally, remark that the star quantization of singular harmonic oscillator, presented here, is different from the Berezin one  discussed in [8,10]. An interesting question concerns a better understanding of the realtionship between the Voros star product using Barut-Girardello states and the noncommutative star product defined in [8] by mean of group-theoretic coherent states. Such relation can be established because the analytical representations of these two sets of coherent states are related through Laplace transforms [23]. Remark also that the deformation quantization discussed in this section can be applied to other quantum mechanical systems (Morse potential, for example).\\
\section{\bf Concluding remarks}
To conclude, let us summarize the main points developed in this work. We have investigated an extension of the standard Voros product using coherent states defined from an arbitrary reference state. Conventional star product is recovered when the reference state coincides with the vacuum. We have shown that the covariance of star calculus can not be guaranteed under all unitary transformations except the canonical ones. Another result obtained in this work concerns the star product derived from the Barut-Girardello coherent states of the $su(1,1)$ algebra. As application, we considered the star analogue of the singular harmonic oscillator. It is interesting to use the extended Voros product in the context of deformation quantization of geometric quantum mechanics purposed in [24] and in the scheme of tomographic representation [25] . It is also desirable to extend the formalism presented here to fermionic and supersymmetric systems as well as other Lie algebras like, for instance, $su(p,q)$. We postpone a full study of such questions to future works.\\
{\vskip 0.1cm}
{\bf Acknowledgements}:
The author would like to thank the Condensed matter Section of the Abdus Salam-ICTP for hospitality. He is also grateful to the referees for theirs comments and suggestions.\\
\vfill\eject

\end{document}